\documentclass[11pt]{article}
\usepackage[dvips]{graphicx}
\usepackage[tbtags]{amsmath}
\usepackage{amsopn}
\usepackage{amsfonts}
\title{Practical non-Abelian Stokes theorem for topologically
	nontrivial Wilson loops}
\author{Bogus\l aw Broda\footnote{
e-mail: bobroda@krysia.uni.lodz.pl}
\, and    Grzegorz Duniec%
\footnote{e-mail: gduniec@merlin.fic.uni.lodz.pl}\\
Department of Theoretical Physics\\
                         University of \L \'od\'z\\  
                           Pomorska 149/153\\ 
                          PL-90-236 \L \'od\'z\\ 
                                       Poland}
\begin{document}
\maketitle
\noindent
\begin{abstract}A practical implementation of the non-Abelian Stokes theorem for topologically nontrivial loops (knots) with possible intersections is proposed.\end{abstract}\section{Introduction}
The Stokes theorem is one of the central points of analysis on manifolds. The formula
\begin{equation*}
\int _{\partial M}\omega =\int _{M}d\omega ,
\end{equation*}
where \(\omega\) is a $(d-1)$-form on the $d$-dimension manifold $M$, is well-known among physicists. The lowest-dimensional version of the Stokes theorem,
\begin{equation}\label{R01}
\oint _{C}A=\int _{S}F,
\end{equation}
where the strength tensor $F=dA$, and  $C=\partial S$, called the (proper) Stokes theorem, is extremely useful in classical electrodynamics (Abelian gauge theory). Eq.\eqref{R01} known as the Abelian Stokes theorem can be generalized to the non-Abelian case [\ref{B01}]. There are several approaches to the non-Abelian Stokes theorem (NAST), and a lot of various aspects of the NAST have been already discussed [\ref{B03}]. One of them, initiated in [\ref{B04}], concerns the NAST for (possibly) topologically nontrivial Wilson loop(s) $C$. In particular three dimensional case 
(${\bf R}^3$),
it may happen that $C$ is knotted (or linked, for multicomponent $C$), and a direct application of the NAST is impossible. In such instances we must invoke a more general procedure.

Formally, we can write the NAST as 
\begin{equation}\label{R02}
P\exp(i\oint _{C}A)=\mathbb{P}\exp(i\int _{S}\mathcal{F})
\end{equation}
where $P$ and $\mathbb{P}$ are appropriately defined orderings, and $\mathcal{F}$ is the "twisted" non-Abelian curvature $F$ of the connection $A$, $F=dA + \frac{1}{2}\lbrack A,A\rbrack $, see [\ref{B01}] or [\ref{B03}] for details. If $C$ bounds a disk $S$, Eq.\eqref{R02} is directly applicable, if not, one can resort to [\ref{B04}], where a version of the NAST for knots and links has been formulated.

 The aim of this paper is two-fold. Firstly, we intend to make the implicit procedure of [\ref{B04}] more explicit. Secondly, we will present a generalization of the NAST allowing intersecting Wilson loops.

\section{The explicit procedure}
The essence of the standard NAST in operator version is a decomposition of the initial loop $C=\partial S$
onto lassos bounding disks of infinitesimal areas. For a star-like $S$ the procedure is straightforward and well-known but for
a topologically non-trivial $C$ the decomposition becomes cumbersome. An elegant solution of the problem
has been proposed in [\ref{B04}], where the authors have found an (implicit) general decomposition 
of $C$ suitable for a direct application of the NAST. The starting point of their analysis is an arbitrary connected 
orientable two-dimensional surface $S_c$ given in a "canonical" form. Since a knot is always a boundary 
of a surface, the so-called Seifert surface $S_s$ (a connected orientable surface), the problem is solved once an
appropriate decomposition for this surface is established. But it is still unclear how one 
can translate the decomposition of [\ref{B04}] for the surface $S_c$ given in a canonical form onto decomposition
of the actual Seifert surface $S_s$. To fill the gap, we will propose a procedure enabling to smoothly pass 
between $S_c$ and $S_s$. 

To begin with, following [\ref{B02}] we are recalling the construction of the 
Seifert surface $S_s$ for a knot $C$. Let us assign $C$ an orientation, and examine its regular projection. 
Near each crossing point, let us delete the over- and under-crossings, and replace them by "short-cut"
arcs. We now have a disjoint collection of closed curves bounding disks, possibly nested. These disks can
be made disjoint by pushing their interiors slightly off the plane. Now, let us connect them together at the 
old crossings with half-twisted strips to from $S_s$. In the case of a multicomponent $C$ (link), we join 
components by tubes, if necessary. 

\begin{figure}[h] \begin{center}
 \includegraphics[height=1.6in,width=5in]{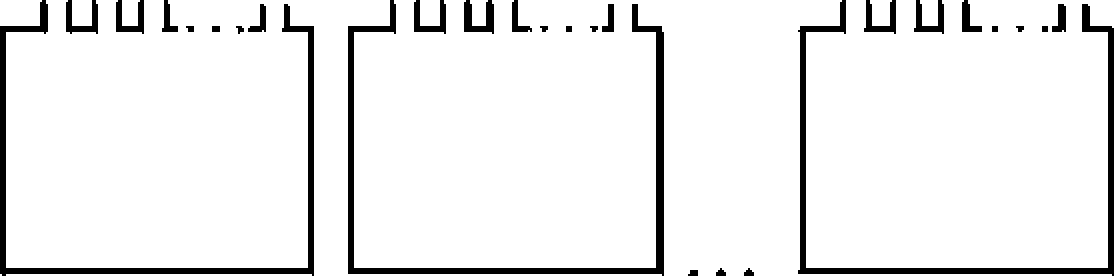}
 \caption{\label{F01}The starting point: ordinary disks connected with strips.}
 \end{center} \end{figure}

\begin{figure}[h] \begin{center}
 \includegraphics[height=2in,width=5in]{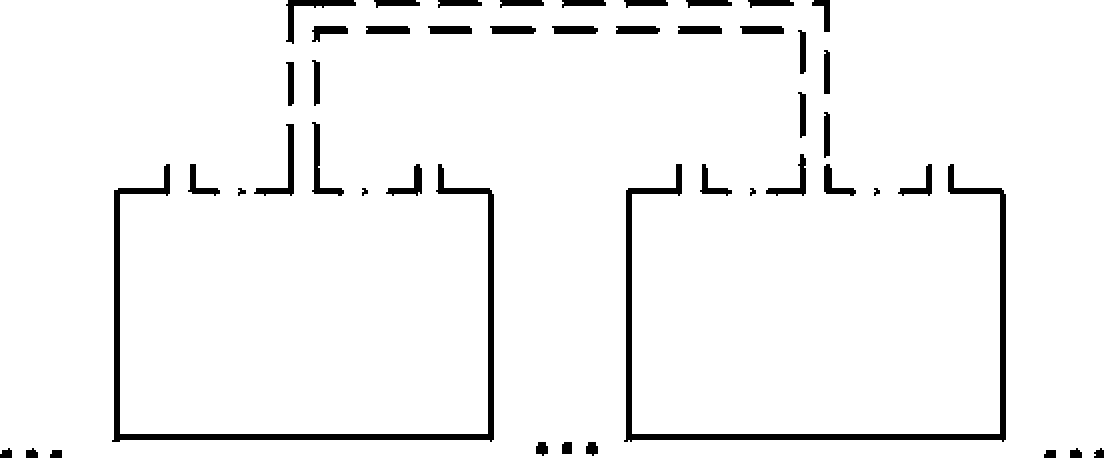}
 \caption{\label{F02} Bringing two connected disks together, and joining them reducing their number by one.}
 \end{center} \end{figure}

\begin{figure}[h]\begin{center}
\includegraphics[height=2in,width=5in]{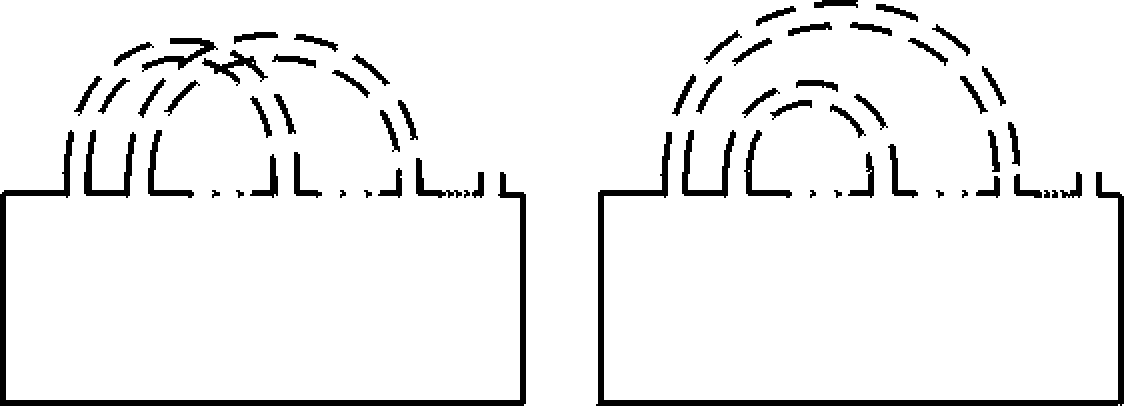}
\caption{\label{F03} The first two strips can be "crossed" or "nested".}
\end{center}\end{figure}

\begin{figure}[h]
\begin{center}
\includegraphics[height=2in,width=5in]{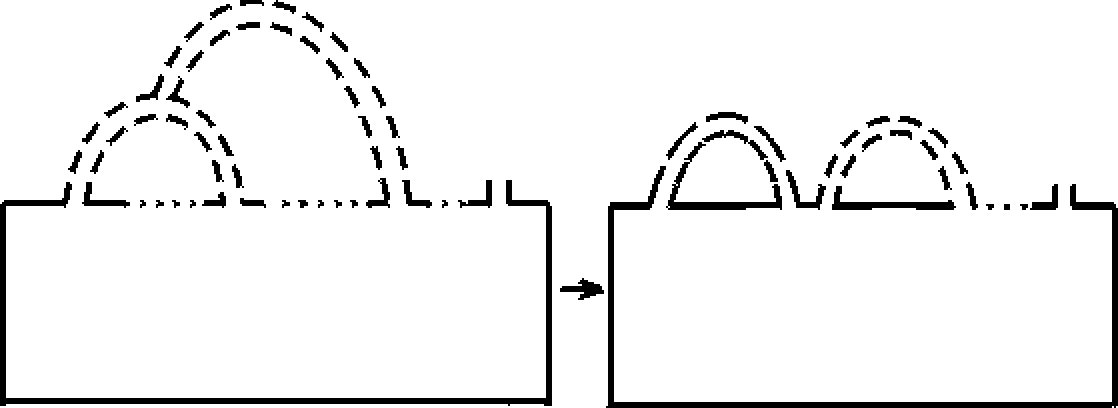}
\caption{\label{F04} Sliding the first strip over the second one we decouple them.}
\end{center}\end{figure}

\begin{figure}[h]\begin{center}
\includegraphics[height=2in,width=5in]{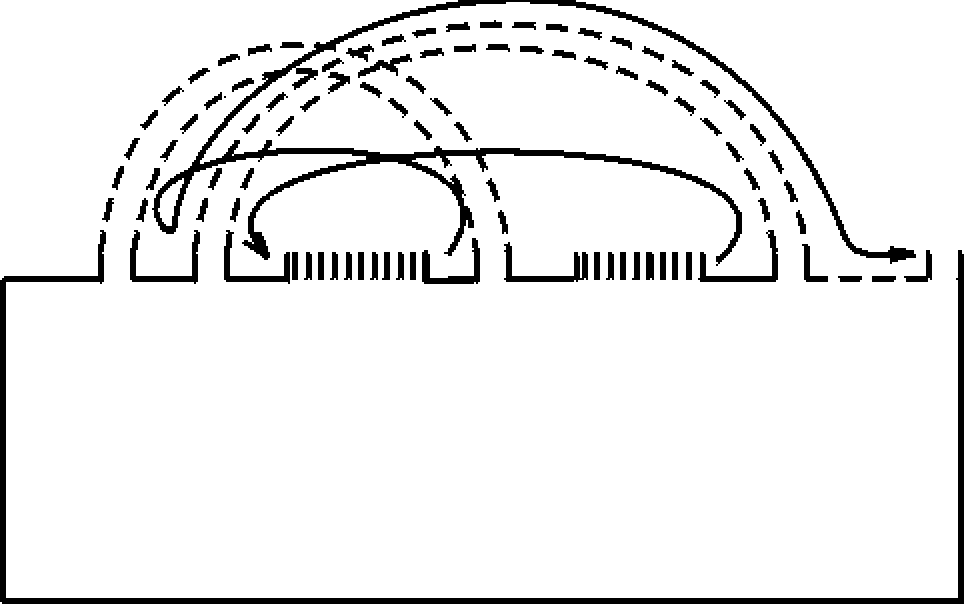}
\caption{\label{F05} Sliding together the whole two bunches of all interior strips ("counter clockwise")
over the two crossed strip separates the bunches from them.}
\end{center}\end{figure}

\begin{figure}[h]
\begin{center}
\includegraphics[height=1.8in,width=5in]{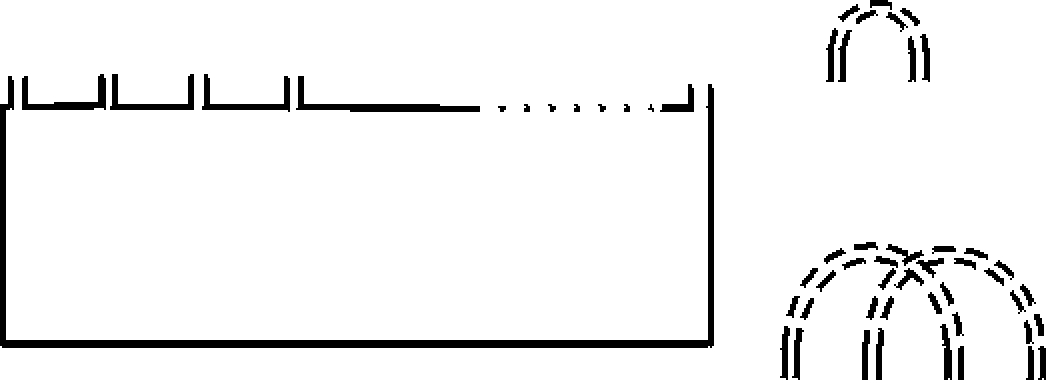}
\caption{\label{F06} We end up with a sequence of "decoupled" single strips and single pairs of crossed strips. We should complete the disk on the left with elements on the right.}
\end{center}\end{figure}

Now, we will describe the procedure of deformation of the Seifert 
surface $S_s$ onto the canonical from $S_c$. To this end, we should realize that according to the previous paragraph our starting point 
are (ordinary)  disks connected with strips (Fig.\ref{F01}).
Shortening a strip, and bringing any two connected disks
together we join them reducing their number by one (Fig.\ref{F02}).
Let us repeat this procedure until we end up with
a single disk. Now, let us concentrate on the first two strips. They can be "crossed" or "nested" (Fig.\ref{F03}).
In the case of nested strips, we can decouple them sliding the first one over the second one, from the left to the right (Fig.\ref{F04}).
In the case of crossed strips, we slide together the whole two bunches of all interior strips ("counter clockwise")  over the two crossed strips separating the bunches from them (Fig.\ref{F05}).
 Let us repeat this procedure until
we end up with a sequence of "decoupled" single strips and single pairs of crossed strips (Fig.\ref{F06}).
Of course, the decoupling takes place only on the boundary of the disk, and the strips can be intertwined in a very complicated way.  

\begin{figure}[h]
\begin{center}
 \includegraphics[height=1.4in,width=5in]{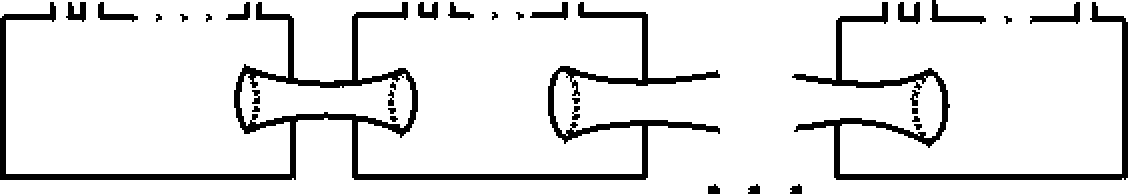}
 \caption{\label{F07} Disconnected disks should be joined by tubes.}
\end{center} \end{figure}

\begin{figure}[h]\begin{center}
\includegraphics[height=1.4in,width=5in]{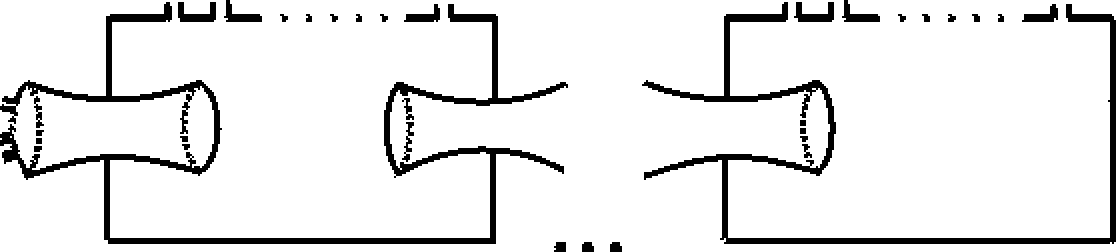}
\caption{\label{F08} Reducing the "size" of the first disk and shortening the first  tube, and next bringing the two first disks together we join them decreasing
their number by one.}
\end{center}\end{figure}

\clearpage
For a link, it may happen that the initial Seifert surface is disconnected. In such a case, after obtaining a single disk for each component, let us join the disks by tubes (Fig.\ref{F07}).
Before we engage in ordering of strips we should cancel the tubes. Reducing the "size" of the first disk and shortening the first  tube, and next bringing the two first disks together we join them decreasing their number by one (Fig.\ref{F08}).
\begin{figure}[h]\begin{center}
\includegraphics[height=1.4in,width=5in]{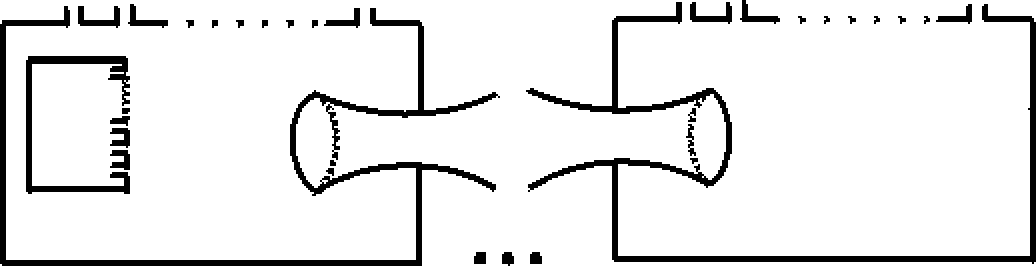}
\caption{\label{F09} Each such on operation creates a hole with strips inside the second disk.}
\end{center}\end{figure}
\begin{figure}[h]
\begin{center}
\includegraphics[height=1.4in,width=5in]{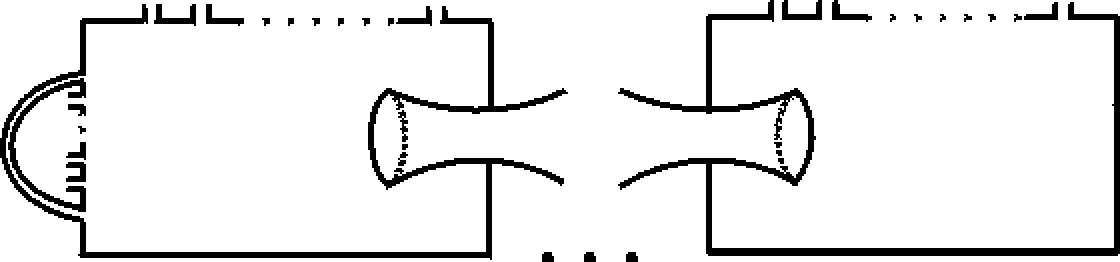}
\caption{\label{F10} Pushing the hole out of the interior of the disk we obtain a standard disk
with a larger number of strips.}
\end{center}\end{figure}
Each such an operation creates o hole with strips
inside the second disk (Fig.\ref{F09}). Pushing the hole out of the interior of the disk we obtain a standard disk with a larger number of strips (Fig.\ref{F10}). Let us keep on repeating the procedure until all the tubes disappear and return to disentangling strips described earlier for a (single) knot $C$. 

\section{Intersections}

\begin{figure}[h]\begin{center}
\includegraphics[height=1.2in,width=3in]{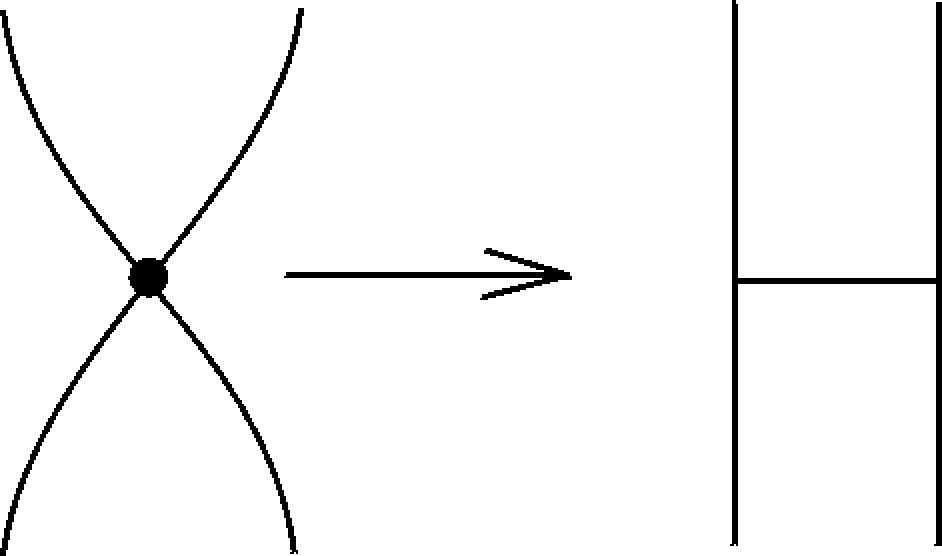}
\caption{\label{F11} Splitting the intersection point.}
\end{center}\end{figure}

The whole method of the previous section can be reused to extend the NAST to the case of 
intersecting Wilson loops (knots/links). Let us return to the construction of the Seifert surface $S_s$. 
Now, some of the crossing points of a regular projection are "true"  crossing points (intersection points).
Splitting the intersection points in an arbitrary way (Fig.\ref{F11})
we get rid off the true crossing points, and the procedure 
of the previous section becomes fully applicable. But the memory about the intersections should remain encoded in the 
form of "pinching" lines identifying the intersection points. These lines lie on  strips, and in the course 
of all necessary arrangements and deformations they persist in lying on the Seifert surface. After joining all the disks all  the lines fall in the 
final disk. The ordering procedure consisting in sliding the strips drags the lines inside the strips. Therefore,
the Seifert surface brought to the canonical form $S_c$ is covered by two independent systems of curves.
The first, very regular one follows from the procedure of [\ref{B04}] and is responsible for cutting $ S_c$ into simply connected surfaces (disks).
The second system of curves, possibly complicated and chaotically looking, generated  by the present procedure, indicates necessary pinching and 
identification of points. A disk created by the first system of curves should be now pinched by curves of the second 
system (Fig.\ref{F11} read backwards) becoming a bunch of disks, and the standard NAST becomes.

\begin{figure}[h]
\begin{center}
\includegraphics[height=3.5in,width=5in]{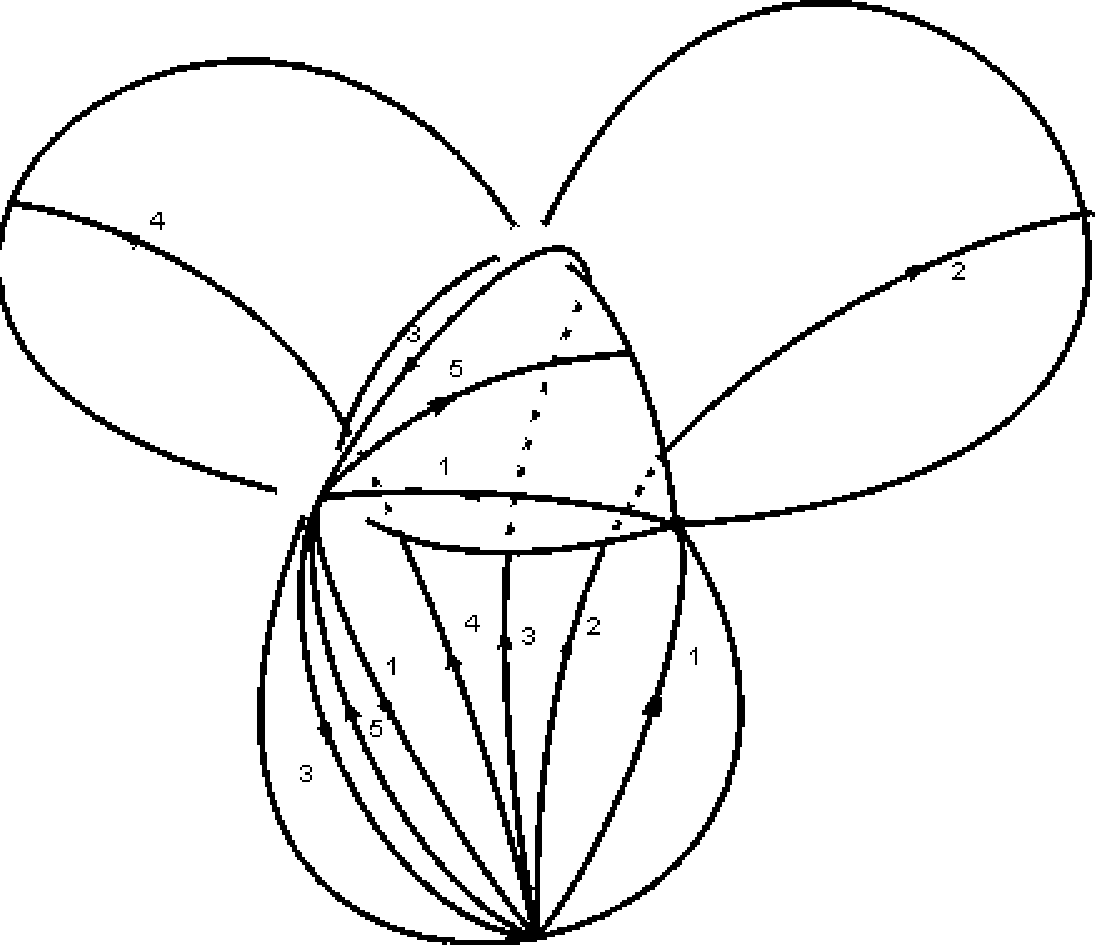}
\caption{\label{F12} The trefoil knot with one intersection: lines without arrows and a dot. Decomposition onto disks: lines with arrows.}\end{center}
\end{figure}

\begin{figure}[h]
\begin{center}\includegraphics[height=5in,width=5in]{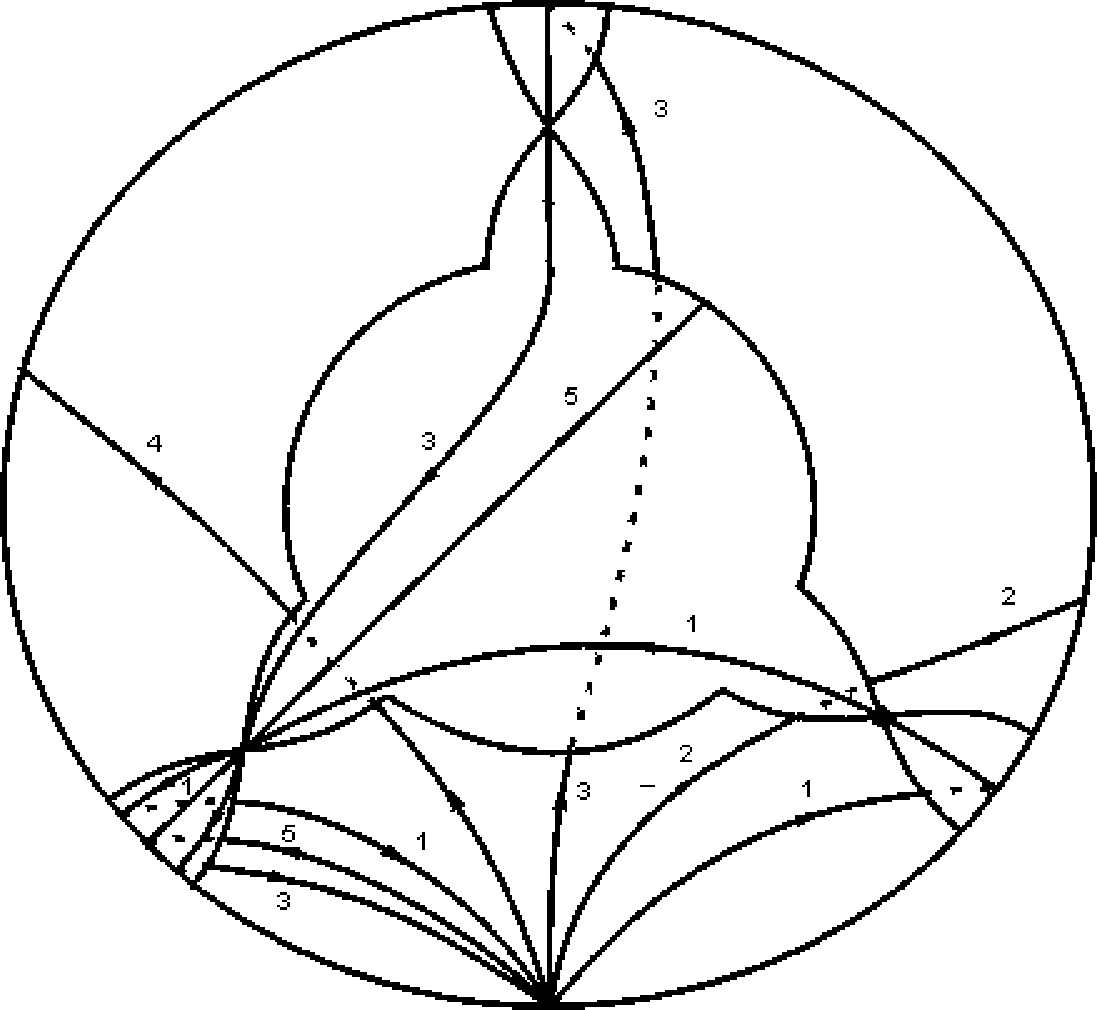}
\caption{\label{F13} An intermediate stage---the Seifert surface $S_s$ with 2 disks.}\end{center}
\end{figure}

\begin{figure}[h]\begin{center}
\includegraphics[height=3.5in,width=5in]{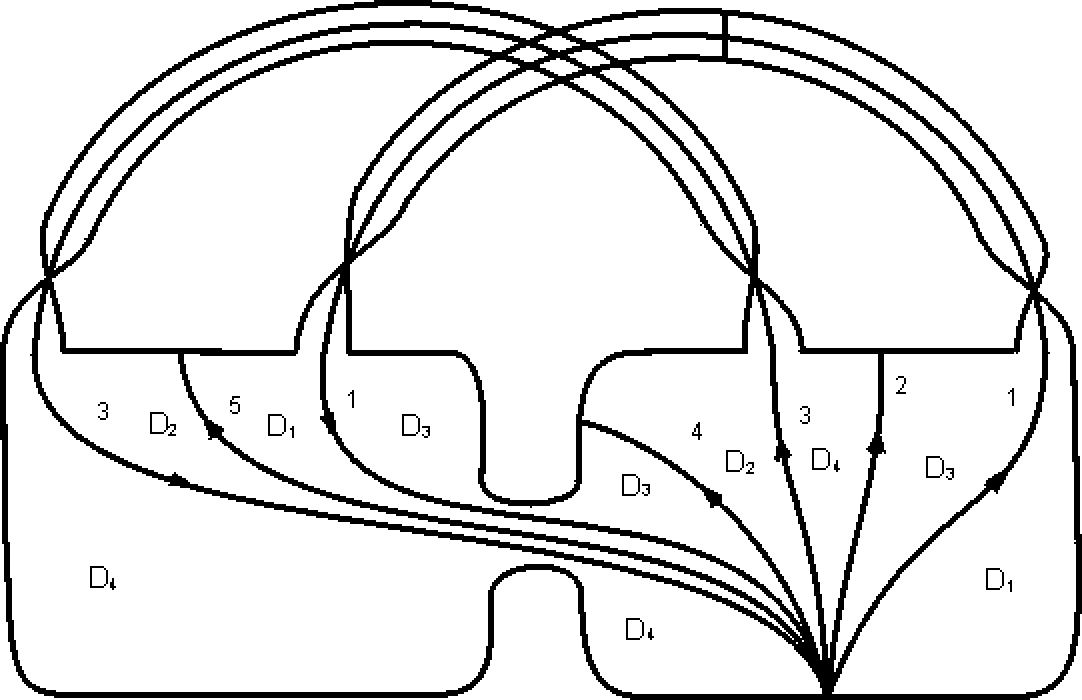}
\caption{\label{F14} An almost canonical form $S_c$.}\end{center}
\end{figure}

\begin{figure}[h]\begin{center}
\includegraphics[height=1.4in,width=5in]{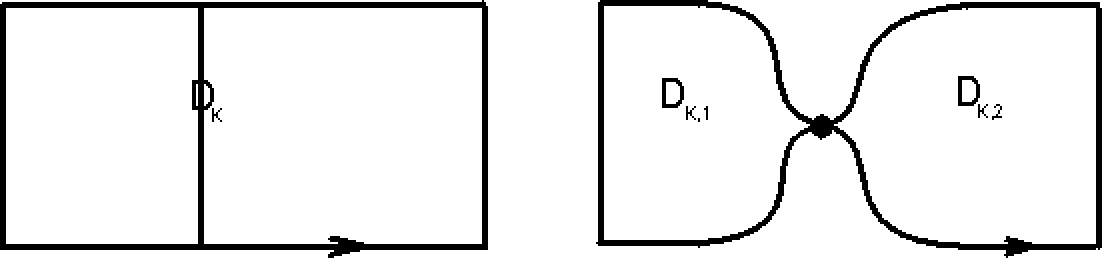}
\caption{\label{F15} The disks $D_1$ and $D_3$ should be cut into two pieces (K=1,3).}
\end{center}
\end{figure}

The best and easiest explanation of the whole procedure can be given in the form of an example.
Let us consider a trefoil knot with one intersection (denoted with a dot) as our example (Fig.\ref{F12}, lines without arrows).
The trefoil can be left- or right-handed, it does not matter.
The primary Seifert surface $S_s$, with 2 disks in this case, is presented in Fig.\ref{F13}, 
whereas its (almost) canonical form $S_c$, in Fig.\ref{F14},  where the pinching line is visible on the right strip.
\clearpage
It follows from Fig.\ref{F14} that exactly the disks $D_1$ and $D_3$ should be cut into two pieces (Fig.\ref{F15}). 
Therefore for example Eq.(2.10) of [\ref{B04}], written is our notation as 
\begin{equation*}
C=[D_a][D_4][D_b]^{-1}[D_3][D_a]^{-1}[D_2][D_b][D_1],
\end{equation*}
should assume now the from
\begin{equation*}
C=[D_a][D_4][D_b]^{-1}[D_{3,1}][D_{3,2}][D_a]^{-1}[D_2][D_b][D_{1,1}][D_{1,2}],
\end{equation*}
where the second subscript counts the newly created, with use of pinching lines, disks. Going backwards from Fig.\ref{F14} to Fig.\ref{F12} we obtain an explicit decomposition of the trefoil knot ready for the application of the standard NAST.
\section{Summary}
In this article, we have shown how the implicit procedure proposed in [\ref{B04}] for the application of the NAST to topologically nontrivial Wilson loops can be practically implemented. As a by-product of our construction, we have extended the theorem to loops with intersections. A general description has been illustrated by an example, i.e. the trefoil knot with one intersection.
\vspace{0.2in}\begin{center}{\bf \large Acknowledgments}
\end{center}The work has been supported by the KBN grant no. 5 P03B 072 21.
\clearpage

\end{document}